\documentclass[revsymb,aps,prb,12pt]{revtex4}
\usepackage{latexsym}
\usepackage[dvips]{graphicx}
\usepackage[cp1251]{inputenc}
\usepackage[english,russian]{babel}
\usepackage{citehack}       
\usepackage{array}

\begin{document}
\draft
\title{Real-space renormalization group study of the anisotropic antiferromagnetic
Heisenberg model of spin $S=1$ on a honeycomb lattice.}
\author{A.S. Boyarchenkov, A.S. Ovchinnikov\cite{byline}, I.G. Bostrem, N.V. Baranov}
\address{Department of Physics, Ural State University, 620083, Ekaterinburg, Russia}
\author{Y. Hosokoshi}
\address{Osaka Prefecture University, Osaka 599-8531, Japan}
\author{K. Inoue}
\address{Department of Chemistry, Hiroshima University, Hiroshima 739-8526, Japan}
\date{\today }

\begin{abstract}
The real-space RG approach is applied to study critical
temperatures of system consisting of interacting spin chains of
spin $S=1$ with an inner antiferromagnetic exchange which form a
honeycomb crystal lattice. Using anisotropic Heisenberg  model we
calculate critical temperature as a function of anisotropic
parameter and the ratio of interchain and intrachain interactions.
A comparison our results with those obtained from RGRS
calculations for the same model of spin-$1/2$ on a square lattice
is given.
\end{abstract}


\maketitle

\section{Introduction.}

The study of low-dimensional magnetic systems made of spin chains attracts
much attention for last decades. If the interchain coupling is less than
intrachain one, at higher temperatures these systems exhibit properties
intrinsic to one-dimensional magnets. However, at lower temperatures the
interchain interaction comes into play and begins to govern a magnetic
behavior of the system.

In this work we apply the real space renormalization group (RSRG) procedure
to the antiferromagnetic Heisenberg (AH)\ model of spin $S=1$ on a honeycomb
lattice made of antiferromagnetic spin $S=1$ chains with an
antiferromagnetic interchain coupling. We are interested in the critical
properties of the 2D system, and explore how the interchain coupling affects
the criticality of the model. Simultaneously, we elucidate an effect of an
exchange anisotropy on the critical temperature and build the phase diagram
of the anisotropic Heisenberg (AH) model of spin $S=1$ on a honeycomb
lattice with two kind of exchange couplings.

First of all, the interest to the problem is motivated by
synthesis of a family of related organic biradicals $PNNNO$,
$PIMNO$ and $F_2PNNNO$ whose magnetic properties have been
examined by susceptibility and magnetization measurements
\cite{Inoue}. Each biradical involves two spins of $S=1/2$, which
are coupled ferromagnetically ($J_F$). In their turn, these spin
pairs couple antiferromagnetically in the crystal. Due to the
strong ferromagnetic coupling $J_F$, antiferromagnetic chains of
$S=1$ are seen in both $PNNNO$ and $F_2PNNNO$ (see Fig. 2(a)
Type-I in Ref.\cite{Inoue}). $PNNNO$ has interchain interactions
in three-dimensions ($3D$), whereas $F_2PNNNO$ has two-dimensional
($2D$) interchain interactions. $PNNNO$ is well understood by
one-dimensional antiferromagnetic chain model. The compound
undergoes $3D$ N\'{e}el order at $\sim $ 1 K due to the weak
interchain coupling. The crystal structures of $F_2PNNNO$ is shown
in Fig.{\ref{structure}}. In the extreme limit of $J_F\to \infty
$, the model becomes equivalent to the coupled
antiferromagnetically ($J_{AF}^{^{\prime }}$) antiferromagnetic
uniform chains of $S=1$ with the intrachain exchange integral $J_{AF}$. $%
F_2PNNNO$ with comparable values of two kinds of the
antiferromagnetic interactions $J_{AF}$ and $J_{AF}^{^{\prime }}$
forms a $2D$ system on a honeycomb lattice. It has been found that
a magnetism of $F_2PNNNO$ is characterized by the singlet ground
state and an energy gap above the state. This finding is supported
by the high-field magnetization process which shows a plateau in
the magnetization curve. We have to point out once that a
modelling of real spin-$1/2$ system by spin $1$ system implies
some mistake and one have to apply the results of the RSRG
analysis to the real compounds with caution.

The results for the spin-1 lattice may be useful also for the theory of $S=1$
bosons (ultracold atoms $^{23}Na$) trapped in an optical lattice in the
regime of one particle per site for suitable interaction between the bosons
\cite{Ho,Yip1}.

According to recent quantum Monte Carlo results for spin-1/2
weakly anisotropic antiferromagnets on the square lattice an
ordered low-temperature phase develops for very small anisotropy
of order $10^{-3} \div 10^{-2}$ (in units of exchange coupling)
\cite{Vaia}. These results are in contradiction with the RSRG\
treatment for the same model predicting considerably larger value
of the critical anisotropy ($\sim 0.2$). In view of this
discrepancy, the RSRG analysis for the $S>1/2$ case is of a
particular theoretical interest and besides  it is instructive  to
compare a critical behavior of systems of integer and half-integer
spins. We mention the results of investigations of ground states of 
spin-$3/2$ (spin-$2$) systems on the hexagonal (square) lattice by 
using vertex state models \cite{Zittartz,Akutsu}. It has been  shown that 
these systems exhibit an anisotropy-induced second order phase transition 
from disordered phase with exponential decay of two-spin correlations 
to the Neel  order phase with dominant long-range correlations. The phase 
transition belongs to the $2D$ Ising-model universality class. 

After the RSRG\ approach has been applied successfully to study
the 2D Ising systems \cite{Niemeijer,Kadanoff1} a number of works
have been dedicated to the investigation of phase transitions in
quantum systems within the method \cite{Rogiers,Stella,Brower}. In
last decade, the RSRG methods have been performed to calculate the
phase diagram for the anisotropic antiferromagnetic Heisenberg
(AAH) model of $S=1/2$ on the square lattice \cite{Souza,Branco}.
This approach uses a hierarchical lattice to approximate the
square one, and performs a partial trace over internal degrees of
freedom.  Recently, in order to study weakly interacting classical
and quantum spin chains the linear-perturbation real-space
renormalization group (LPRG) method has been suggested
\cite{Sznajd}. The LPRG uses the existence of the small parameter
-- the ratio of interchain to intrachain coupling. This
perturbative method involves RG transformation, which for the
Ising spins is the standard decimation procedure and for the
quantum spins is the generalization of the Suzuki-Takano
approximate decimation \cite{Suzuki}. However, in practice, the
LPRG using perturbation theory with the interchain interactions as
the perturbation parameters is only reliable for small values of
the ratio interchain to intrachain interaction.

We use an extension to $S=1$ case of the quantum RSRG\ approach originally
suggested by Mariz et al. for the $S=1/2$ AH model \cite{Mariz}. An
application of the renormalization group\ method to interacting quantum spin
chains encounters standard difficulties connected with a necessity of a
decomposition of the exponential operators, and additional proliferation of
the new interactions due to the vector character of the spin operators.

The paper is organized as follows. In Sec. II\ the RSRG method  is
formulated for $S=1$ and applied to the honeycomb lattice formed
by interacting antiferromagnetic chains. Our results and
conclusions are given in Sec. III.

\section{Model.}

We consider a system (anisotropic Heisenberg model) whose dimensionless
Hamiltonian is defined by
\begin{equation}
-\beta H=\sum\limits_{<i,j>}K_{ij}\left[ (1-\Delta _{ij})\left(
S_i^xS_j^x+S_i^yS_j^y\right) +S_i^zS_j^z\right]   \label{Hamil}
\end{equation}
where $\beta =1/k_BT$, $K_{ij}\equiv J_{ij}/k_BT$, $J_{ij}$ is the exchange
coupling constant, $\left\langle ij\right\rangle $ denotes first-neighboring
lattice sites, $\Delta _{ij}$ is the anisotropic parameter, and $S_i^\alpha $
$\left\{ \alpha =x,y,z\right\} $ is the spin $1$ on the site $i$. The
Hamiltonian (\ref{Hamil}) describes the Ising ($\Delta _{ij}=1$), isotropic
Heisenberg ($\Delta _{ij}=0$) and XY-model ($\Delta _{ij}=-\infty $).

We mention briefly main operations lying in base of the approach of Ref.
\cite{Mariz}. A parallel array of two bonds characterized respectively by $%
(K_1,\Delta _1)$ and $(K_2,\Delta _2)$ is equivalent to a single bond
characterized by $(K_p,\Delta _p)$ given by
\[
K_p=K_1+K_2,
\]
\[
K_p\Delta _p=K_1\Delta _1+K_2\Delta _2.
\]
The extension of the approach, that is nothing but Migdal-Kadanoff procedure
\cite{Migdal,Kadanoff}, to $n$ parallel bonds is straightforward.

For a series array of two bonds characterized by $(K_1,\Delta _1)$ and $%
(K_2,\Delta _2)$ the procedure involves difficulties due to
non-commutativity effects. Under rescaling and removal of intermediate spins
(decimation), the Hamiltonian changes and is characterized by a new set of
parameters that are functions of the original set. The initial Hamiltonian
is given by
\[
H_{123}=K_1\left[ \left( 1-\Delta _1\right) \left(
S_1^xS_3^x+S_1^yS_3^y\right) +S_1^zS_3^z\right] +K_2\left[ \left( 1-\Delta
_2\right) \left( S_3^xS_2^x+S_3^yS_2^y\right) +S_3^zS_2^z\right] .
\]
We have to replace this array by a single bond whose Hamiltonian is
\begin{equation}
H_{12}^{\prime }=K_s\left[ \left( 1-\Delta _s\right) \left(
S_1^xS_2^x+S_1^yS_2^y\right) +S_1^zS_2^z\right] +K_0^{\prime }.
\label{H12eff}
\end{equation}
We impose the preservation of the partition function between terminal sites $%
1$ and $2$, i.e.
\begin{equation}
\exp H_{12}^{\prime }=\text{Tr}_3\exp H_{132},  \label{RGs}
\end{equation}
where Tr$_3$ denotes the tracing operation over the states of the
intermediate spin $3$, $K_0^{\prime }$ is an additive constant included to
make Eq. (\ref{RGs}) possible. Equation (\ref{RGs}) establishes the relation
between the set of parameters $(K_1,\Delta _1)$, $(K_2,\Delta _2)$ and the
set of renormalized parameters ($K_s,\Delta _s,K_0^{\prime }$). For the
anisotropic spin-$1/2$ Heisenberg model corresponding expressions may be
found explicitly \cite{Mariz},\cite{Souza}.

In order to construct recursion relations of the RG\ transformation (\ref
{RGs}) for the spin-$1$ model we should expand both sides over an
appropriate matrix basis to equate coefficients of the expansion. To avoid a
proliferation problem we chose those quantities that correspond to coupling
constants in the initial Hamiltonian (\ref{Hamil}).

Due to properties of Pauli matrices the matrix representation for $H$ and $%
\exp H$ in the case of spin $1/2$ has the same form, i.e. $H$ and $\exp H$
have non-zero matrix elements in the same positions. For the $S=1$ this rule
does not hold that results in an essential complication of RG procedure.

Following the treatment of Ref. \cite{Mariz}, we expand $\exp H_{12}^{\prime
}$ as

\[
\exp H_{12}^{\prime }=\sum\limits_{n_1=0}^\infty \sum\limits_{n_2=0}^\infty
K_{n_1n_2}A_1^{n_1}\otimes A_2^{n_2},
\]
where $\otimes $ is the outer product, $A_{1,2}$ is ordinary
powers of the spin operators $S_{1,2}^{x,y,z}$ and the
coefficients $K_{n_1n_2}$
depend on $K_s,\Delta _s$ and $K_0^{\prime }$. Since, $A_{1,2}^n$ are the $%
3\times 3$ matrices we expand them over the basis that consists of the
polarization matrices $T_q^k$ ($k=0,1,2$ and $q=-k,-k+1,\ldots ,k$) (see
Appendix A)
\[
A_i^n=a\left( T_i\right) _0^0+\sum\limits_{M=\pm 1,0}b^M\left( T_i\right)
_M^1+\sum\limits_{M=\pm 2,\pm 1,0}c^M\left( T_i\right) _M^2,\;(i=1,2).
\]
In their turn, the matrices $T_q^k$ may be written through the spin
operators explicitly \cite{Varshalovich}
\[
T_0^0=\frac 1{\sqrt{3}}I,\;T_{\pm 1}^1=\mp \frac 12\left( S_x\pm iS_y\right)
,\;T_0^2=\sqrt{\frac 32}\left( \left( S^z\right) ^2-\frac 23I\right) ,
\]
\[
T_{\pm 1}^2=\mp \frac 12\left[ \left( S_xS_z+S_zS_x\right) \pm i\left(
S_yS_z+S_zS_y\right) \right] ,\;T_{\pm 2}^2=\frac 12\left[ \left( S^x\right)
^2-\left( S^y\right) ^2\pm i\left( S_xS_y+S_yS_x\right) \right] .
\]
The transformation $\exp (H)\,$ should preserve the symmetry of the
Hamiltonian $H$. Thus, the requirement of invariance gives the most general
form
\[
\exp H_{12}=\alpha _1\left( \left( T_1\right) _0^0\otimes \left( T_2\right)
_0^0\right) +\alpha _2\left( \left( T_1\right) _0^1\otimes \left( T_2\right)
_0^1\right) +\alpha _3\left( \left( T_1\right) _0^2\otimes \left( T_2\right)
_0^2\right) +\beta \left( \left( T_1\right) _1^1\otimes \left( T_2\right)
_{-1}^1+\left( T_1\right) _{-1}^1\otimes \left( T_2\right) _1^1\right)
\]
\begin{equation}
+\gamma \left( \left( T_1\right) _2^2\otimes \left( T_2\right)
_{-2}^2+\left( T_1\right) _{-2}^2\otimes \left( T_2\right) _2^2\right)
+\sigma \left( \left( T_1\right) _1^2\otimes \left( T_2\right)
_{-1}^2+\left( T_1\right) _{-1}^2\otimes \left( T_2\right) _1^2\right)
\label{expan}
\end{equation}
with a new set of coupling parameters $\alpha _1$, $\alpha _2$, $\alpha _3$,
$\beta $, $\gamma $, $\sigma $. For the Hamiltonian (\ref{H12eff}) the above
formula gives
\begin{equation}
\exp H_{12}=\left(
\begin{array}{ccccccccc}
A_{12} & 0 & 0 & 0 & 0 & 0 & 0 & 0 & 0 \\
0 & B_{12} & 0 & C_{12} & 0 & 0 & 0 & 0 & 0 \\
0 & 0 & D_{12} & 0 & F_{12} & 0 & G_{12} & 0 & 0 \\
0 & C_{12} & 0 & B_{12} & 0 & 0 & 0 & 0 & 0 \\
0 & 0 & F_{12} & 0 & E_{12} & 0 & F_{12} & 0 & 0 \\
0 & 0 & 0 & 0 & 0 & B_{12} & 0 & C_{12} & 0 \\
0 & 0 & G_{12} & 0 & F_{12} & 0 & D_{12} & 0 & 0 \\
0 & 0 & 0 & 0 & 0 & C & 0 & B_{12} & 0 \\
0 & 0 & 0 & 0 & 0 & 0 & 0 & 0 & A_{12}
\end{array}
\right) ,  \label{struct}
\end{equation}
where the matrix elements are $A_{12}\equiv \alpha _1/3+\alpha _2/2+\alpha
_3/6$, $B_{12}\equiv \alpha _1/3-\alpha _3/3$, $C_{12}\equiv -\beta
/2-\sigma /2$, $D_{12}\equiv \alpha _1/3-\alpha _2/2+\alpha _3/6$, $%
E_{12}\equiv \alpha _1/3+2\alpha _3/3,\;F_{12}\equiv -\beta /2+\sigma
/2,\;G_{12}\equiv \gamma .$

Similarly, we obtain the closed expression for the expansion of $Tr_3\exp
H_{123}$ akin to that of $\exp H_{12}$ with the coefficients in the
expansion (\ref{expan}) are functions of the parameters coming into $H_{123}$
(see Appendix B for details).

To calculate the exponentials we will diagonalize numerically the $9\times
9\,$ and $27\times 27$ matrices associated with $\,H_{12}$ and $H_{123}$. By
using
\[
\exp H_{12}=U_{12}\exp (H_{12}^D)U_{12}^{\dagger },\;\exp
H_{123}=U_{123}\exp (H_{123}^D)U_{123}^{\dagger },
\]
where $U_{12},U_{123}$ are the unitary matrices turning $H_{12},H_{123}$
into the diagonal forms $H_{12}^D,H_{123}^D$, we can find numerically $\exp
H_{12}$ and $\exp H_{123}$ as functions of corresponding coupling
parameters. The same matrix structure (\ref{struct}) of $\exp H_{12}$ and $%
Tr_3\exp H_{123}$ is supported by the numerical calculation. The numerical
procedure produces a set $\left\{ \alpha _1,\alpha _2,\alpha _3,\beta
,\gamma ,\sigma \right\} $ for $\exp H_{12}$ and $\left\{ \bar{\alpha}_1,%
\bar{\alpha}_2,\bar{\alpha}_3,\bar{\beta},\bar{\gamma},\bar{\sigma}\right\} $
for Tr$_3\exp H_{132}$. To obtain the required RG equations we impose

\begin{equation}
\alpha _1=A_{12}+B_{12}+D_{12}=\bar{A}_{12}+\bar{B}_{12}+\bar{D}_{12}=\bar{%
\alpha}_1,  \label{rg1}
\end{equation}
\begin{equation}
\alpha _2=A_{12}-D_{12}=\bar{A}_{12}-\bar{D}_{12}=\bar{\alpha}_2,
\label{rg2}
\end{equation}
\begin{equation}
\beta =-C_{12}-F_{12}=-\bar{C}_{12}-\bar{F}_{12}=\bar{\beta}  \label{rg3}
\end{equation}
and
\begin{equation}
\alpha _3=E_{12}-B_{12}=\bar{E}_{12}-\bar{B}_{12}=\bar{\alpha}_3,\,\;\gamma
=G_{12}=\bar{G}_{12}=\bar{\gamma},\,\;\sigma =F_{12}-C_{12}=\bar{F}_{12}-%
\bar{C}_{12}=\bar{\sigma}.  \label{rg4}
\end{equation}
The number of these equations exceeds the number of interactions that enter
into the Hamiltonian (\ref{H12eff}) because all possible bilinear couplings
between terminal sites come into play. ~Thus, in order to carry out the RG\
decimation we retain the three equations (\ref{rg1},\ref{rg2},\ref{rg3})
which implicitly determine $K_s,\Delta _s$ and $K_0^{\prime }$ as functions
of $(K_1,\Delta _1)$, $(K_2,\Delta _2)$. This set of equations is a
counterpart of RG relations for the case of $S=1/2$ (see Eqs.(12) in Ref.
\cite{Mariz}).

Now, we have to chose an appropriate hierarchical lattice. We take
one of the simplest cells, conserving a point symmetry of the full
lattice, with $6$ sites and $6$ bonds, depicted in
Fig.{\ref{graph}}. We then impose that the correlation function
between the two terminal sites $3$ and $6$ of the original and
renormalized graphs are preserved. At the first step we apply
decimation, the spins $1$ and $3$ (or $4$ and $6$) survive whereas
the spins $2$ and $5$ are removed. At the second step the
decimation procedure is repeated removing the spins $1$ and $4$.
Finally, to obtain the renormalized parameters we apply
Migdal-Kadanoff bond moving combining the ''pieces'' in parallel,
which leads to the recursion relations
\begin{equation}
\begin{array}{c}
\left( K_S,\Delta _S\right) =R_S\left( K_2,\Delta _2;K_1,\Delta _1\right) ,
\\
\left( K_S^{^{\prime }},\Delta _S^{^{\prime }}\right) =R_S\left( K_S,\Delta
_S;K_1,\Delta _1\right) , \\
\left( K_p,\Delta _p\right) =2\left( K_S^{^{\prime }},\Delta _S^{^{\prime
}}\right) .
\end{array}
\label{RGeq}
\end{equation}

We have evaluated numerically the renormalization transformation from the
original set of coupling parameters to the set of renormalized parameters.
Critical points are then evaluated as non-trivial fixed points of the above
relations which can be rewritten as the composite function
\begin{equation}
\left( K_p,\Delta _p\right) =2R_S\left( R_S\left( K_1,\Delta _1;K_2,\Delta
_2\right) ;K_1,\Delta _1\right) \text{.}  \label{composite}
\end{equation}

Unlike to the case of $S=1/2$ we can not obtain RG\ relations explicitly.
Instead of this, we briefly outline the numerical procedure. Input fixed
parameters are the ratio of interchain to intrachain coupling $%
C_1=J_{AF}^{^{\prime }}/J_{AF}$ and the anisotropic parameter
$\Delta =\Delta _1$ and what's more $\Delta _2=C_1\Delta $ (the
feature of anisotropy is the same both
for the intrachain and interchain couplings). At given starting value $%
K_{1i}$ of the intrachain coupling (then the interchain coupling is $%
K_{2i}=C_1K_{1i}$) we apply two successive decimation steps to produce a
renormalized coupling $K_S^{^{\prime }}$. During each of the step we solve
Eqs.(\ref{rg1},\ref{rg2},\ref{rg3}) using the standard routine for
non-linear systems of equations \cite{Press}. Then, we double the result
obtained after these transformations to get a final value $K_f$ depending on
the starting value $K_{1i}$. To complete we find a fixed point $K_c$ of the
equation $K_{1i}=K_f(K_{1i})$ by using bisection method.

\section{Results.}

The critical inverse temperature $K_c=1/T_c$ as a function of $C_1$ for
several $\Delta $ values is presented in Fig.{\ref{critcoupl}}. As seen, the
critical temperature rapidly decreases when the interchain coupling becomes
weaker. Our results for the critical temperature as a function of the
anisotropy $\Delta $ are shown in Fig.{\ref{critanis}}. The universality
class for the whole critical curve is the same as for the Ising model. By
contrast to some RSRG calculations for $S=1/2$ the phase diagrams for
ferromagnetic and antiferromagnetic models are the same: the critical
temperature reaches zero at a critical value of $\Delta $, $\Delta _c$,
which is greater than zero. The weaker interchain coupling the stronger
quantum fluctuations. So, as one could expect, the $\Delta _c$ value is
larger if the $C_1$ shifts to lower values.

For the lowest temperatures we could work we have observed no sign of the
reentrant behavior found in some previous RGRS treatments. The N\'{e}el
temperature behaves as
\[
T_N\sim \frac 1{\ln \left( \Delta -\Delta _c\right) }
\]
near $\Delta =\Delta _c$ that agrees with the result found for the case $%
S=1/2$. Our calculation can not be carried out down to $T=0$;
therefore we can not make any definite conclusions about the
ground state of the model. The scaling law holds for different
values of the ratio of interchain to intrachain coupling
(Fig.{\ref{scaling}}). In Ref. \cite{Ding,Aplesnin}, the
logarithmic dependence of $T_N$ and $T_c$ with respect to $\Delta
-\Delta _c$ is established using scaling arguments with $\Delta
_c=0$. Recent quantum Monte Carlo results  \cite{Vaia} for the
anisotropic 2D $S=1/2$ Heisenberg model have shown that it
develops an ordered low-temperature phase even for very small
anisotropies $\Delta \sim 10^{-3},10^{-2}$. The latter gives
strong evidence of large values of a critical anisotropy is an
artifact of the real-space renormalization approach.

At this point, it is worthwhile to compare our results with those
from RGRS calculations for antiferromagnetic  AH model of $S=1/2$
on a square lattice. These RSRG analysises lead to non-equivalence
between the criticality of the ferromagnetic (F) and
antiferromagnetic (AF) models, a reentrant behavior in the ($T$,
$\Delta $) diagram \cite{Souza,Branco}.

It is well known that in classical spin models, such as as the
Ising or classical Heisenberg models, on bipartite lattices the
critical temperature is the same for ferromagnetic exchange (Curie
temperature) as for antiferromagnetic exchange (Neel temperature).
This is a direct consequence of the free energy being an even
function of the exchange parameter. However, for the quantum spin
$1/2$ Heisenberg model the Curie and Neel temperatures are unequal
\cite{Rushbrooke}. Recently, this question has been reinvestigated
using high-temperature series expansions for the spin $1/2$, $1 $
and $3/2$ Heisenberg ferromagnet and antiferromagnet in
$3$-dimensions \cite{Oitmaa}. The difference between the
temperatures decreases rapidly
with increasing spin $S$. In some quantum systems, such as the quantum spin $%
1/2$ XY and transverse Ising models, an isomorphism between the criticality
of the F and AF cases is observed \cite{SousaJMMM}. Critical properties of
the quantum spin-1/2 2D Heisenberg model with anisotropic interaction
treated by the Green's function technique yields $T_c=T_N$ for all values of
the anisotropy parameter \cite{Singh,Liu}. RSRG\ methods give contradictory
results of the problem because of underlying approximations whose effects
are hard to control in a systematic way. In Refs. \cite{Souza,Branco} it was
obtained by RG approach $T_N<T_c$ for 2D anisotropic Heisenberg limit $0\leq
\Delta <1$ due to a special choice of the hierarchical lattice approximating
a square one. The critical temperature $T_c$ for 2D ferromagnetic spin $1/2$
anisotropic Heisenberg model tends gradually to zero when decreasing the
anisotropy parameter $\Delta $, i.e. $T_c=0$ in the isotropic Heisenberg
limit $\Delta =0$ in accordance with Mermin-Wagner theorem \cite{Mermin}.
The results for the antiferromagnetic exchange are very similar to the $S=1/2
$ case, there is no long-range N\'{e}el order for the anisotropy parameter $%
\Delta <{\Delta }_c$. The question is not yet settled and more work is
needed to put these points on firmer grounds.

An observation of reentrance behavior for the spin 1/2
antiferromagnetic AH model on the square lattice has been reported
by some authors \cite{Souza,Branco}. This result suggests that
there is an ordered phase at relatively high temperature but not
at very low temperature. A full understanding of this phenomena is
still lacking, but it's most likely the reentrance behavior is an
artifact of the RGRS method. In Refs. \cite{Araujo} the authors
attribute the reentrance to the effect of finite size in the
renormalization and for larger clusters it should be absent.

In conclusion, the real-space renormalization group  is employed
to study the anisotropic Heisenberg model of spin $S=1$ on a
honeycomb lattice with two kinds of antiferromagnetic couplings.
We calculate dependencies of the critical temperature on the
parameters of the magnetic anisotropy, and on the ratio of
interchain and intrachain exchange interactions. The entire
critical line is found to belong to the universality class of the
Ising model. In accordance with the early  RSRG predictions for
the antiferromagnetic AH model of spin $S=1/2$ on a square lattice
our
calculations recover an existence of large finite critical anisotropy ${%
\Delta }_c$   that should be considered, however, as an artifact
of the real-space renormalization technique.

\acknowledgments
We would like to thank Prof. M. V. Sadovskii for discussions. We
acknowledges partial financial support from the grant NREC-005 of US CRDF
(Civilian Research \& Development Foundation).

\section*{Appendix A}

The polarization matrices for the spin $S=1$ has the explicit form

\[
T_{00}=\frac 13\left(
\begin{array}{ccc}
1 & 0 & 0 \\
0 & 1 & 0 \\
0 & 0 & 1
\end{array}
\right) ,\;T_{11}=-\frac 1{\sqrt{2}}\left(
\begin{array}{ccc}
0 & 1 & 0 \\
0 & 0 & 1 \\
0 & 0 & 0
\end{array}
\right) ,\;T_{1-1}=\frac 1{\sqrt{2}}\left(
\begin{array}{ccc}
0 & 0 & 0 \\
1 & 0 & 0 \\
0 & 1 & 0
\end{array}
\right) ,
\]
\[
T_{10}=\frac 1{\sqrt{2}}\left(
\begin{array}{ccc}
1 & 0 & 0 \\
0 & 0 & 0 \\
0 & 0 & -1
\end{array}
\right) ,\;T_{20}=\frac 1{\sqrt{6}}\left(
\begin{array}{ccc}
1 & 0 & 0 \\
0 & -2 & 0 \\
0 & 0 & 1
\end{array}
\right) ,\;T_{21}=\frac 1{\sqrt{2}}\left(
\begin{array}{ccc}
0 & -1 & 0 \\
0 & 0 & 1 \\
0 & 0 & 0
\end{array}
\right) ,
\]
\[
T_{2-1}=\frac 1{\sqrt{2}}\left(
\begin{array}{ccc}
0 & 0 & 0 \\
1 & 0 & 0 \\
0 & -1 & 0
\end{array}
\right) ,\;T_{22}=\left(
\begin{array}{ccc}
0 & 0 & 1 \\
0 & 0 & 0 \\
0 & 0 & 0
\end{array}
\right) ,\;T_{2-2}=\left(
\begin{array}{ccc}
0 & 0 & 0 \\
0 & 0 & 0 \\
1 & 0 & 0
\end{array}
\right) .
\]

\section*{Appendix B}

We first take care of $H_{12}^{^{\prime }}$ and express it in the basis $%
\left| M_1M_2\right\rangle $. In this basis $H_{12}^{^{\prime }}$ becomes

\[
H_{12}^{^{\prime }}=\left(
\begin{array}{ccccccccc}
K_0^{^{\prime }}+K_S & 0 & 0 & 0 & 0 & 0 & 0 & 0 & 0 \\
0 & K_0^{^{\prime }} & 0 & W_S & 0 & 0 & 0 & 0 & 0 \\
0 & 0 & K_0^{^{\prime }}-K_S & 0 & W_S & 0 & 0 & 0 & 0 \\
0 & W_S & 0 & K_0^{^{\prime }} & 0 & 0 & 0 & 0 & 0 \\
0 & 0 & W_S & 0 & K_0^{^{\prime }} & 0 & W_S & 0 & 0 \\
0 & 0 & 0 & 0 & 0 & K_0^{^{\prime }} & 0 & W_S & 0 \\
0 & 0 & 0 & 0 & W_S & 0 & K_0^{^{\prime }}-K_S & 0 & 0 \\
0 & 0 & 0 & 0 & 0 & W_S & 0 & K_0^{^{\prime }} & 0 \\
0 & 0 & 0 & 0 & 0 & 0 & 0 & 0 & K_0^{^{\prime }}+K_S
\end{array}
\right)
\]
where $W_S=K_S\left( 1-\Delta _S\right) $.

Performing the same calculation for $H_{123}$, now using the basis $\left|
M_1M_3M_2\right\rangle $, we arrive at a $27\times 27$ which has $4$
independent blocks of size $9\times 9$
\[
H_{123}=\left(
\begin{array}{ccc}
A_1 & B_2 & 0 \\
B_1 & A_2 & B_2 \\
0 & B_1 & A_3
\end{array}
\right) ,
\]
where the $9\times 9$ $A_1$, $B_1$, $B_2$, $A_2$ and $A_3$ are given by
\[
A_1=\left(
\begin{array}{ccccccccc}
K_1+K_2 & 0 & 0 & 0 & 0 & 0 & 0 & 0 & 0 \\
0 & K_1 & 0 & W_2 & 0 & 0 & 0 & 0 & 0 \\
0 & 0 & K_1-K_2 & 0 & W_2 & 0 & 0 & 0 & 0 \\
0 & W_2 & 0 & 0 & 0 & 0 & 0 & 0 & 0 \\
0 & 0 & W_2 & 0 & 0 & 0 & W_2 & 0 & 0 \\
0 & 0 & 0 & 0 & 0 & 0 & 0 & W_2 & 0 \\
0 & 0 & 0 & 0 & W_2 & 0 & -K_1-K_2 & 0 & 0 \\
0 & 0 & 0 & 0 & 0 & W_2 & 0 & -K_1 & 0 \\
0 & 0 & 0 & 0 & 0 & 0 & 0 & 0 & -K_1+K_2
\end{array}
\right) ,
\]
\[
A_2=\left(
\begin{array}{ccccccccc}
K_2 & 0 & 0 & 0 & 0 & 0 & 0 & 0 & 0 \\
0 & 0 & 0 & W_2 & 0 & 0 & 0 & 0 & 0 \\
0 & 0 & -K_2 & 0 & W_2 & 0 & 0 & 0 & 0 \\
0 & W_2 & 0 & 0 & 0 & 0 & 0 & 0 & 0 \\
0 & 0 & W_2 & 0 & 0 & 0 & W_2 & 0 & 0 \\
0 & 0 & 0 & 0 & 0 & 0 & 0 & W_2 & 0 \\
0 & 0 & 0 & 0 & W_2 & 0 & -K_2 & 0 & 0 \\
0 & 0 & 0 & 0 & 0 & W_2 & 0 & 0 & 0 \\
0 & 0 & 0 & 0 & 0 & 0 & 0 & 0 & K_2
\end{array}
\right) ,
\]
\[
A_3=\left(
\begin{array}{ccccccccc}
-K_1+K_2 & 0 & 0 & 0 & 0 & 0 & 0 & 0 & 0 \\
0 & -K_1 & 0 & W_2 & 0 & 0 & 0 & 0 & 0 \\
0 & 0 & -K_1-K_2 & 0 & W_2 & 0 & 0 & 0 & 0 \\
0 & W_2 & 0 & 0 & 0 & 0 & 0 & 0 & 0 \\
0 & 0 & W_2 & 0 & 0 & 0 & W_2 & 0 & 0 \\
0 & 0 & 0 & 0 & 0 & 0 & 0 & W_2 & 0 \\
0 & 0 & 0 & 0 & W_2 & 0 & K_1-K_2 & 0 & 0 \\
0 & 0 & 0 & 0 & 0 & W_2 & 0 & K_1 & 0 \\
0 & 0 & 0 & 0 & 0 & 0 & 0 & 0 & K_1+K_2
\end{array}
\right) ,
\]
\[
B_2=B_1^T=\left(
\begin{array}{ccccccccc}
0 & 0 & 0 & 0 & 0 & 0 & 0 & 0 & 0 \\
0 & 0 & 0 & 0 & 0 & 0 & 0 & 0 & 0 \\
0 & 0 & 0 & 0 & 0 & 0 & 0 & 0 & 0 \\
W_1 & 0 & 0 & 0 & 0 & 0 & 0 & 0 & 0 \\
0 & W_1 & 0 & 0 & 0 & 0 & 0 & 0 & 0 \\
0 & 0 & W_1 & 0 & 0 & 0 & 0 & 0 & 0 \\
0 & 0 & 0 & W_1 & 0 & 0 & 0 & 0 & 0 \\
0 & 0 & 0 & 0 & W_1 & 0 & 0 & 0 & 0 \\
0 & 0 & 0 & 0 & 0 & W_1 & 0 & 0 & 0
\end{array}
\right) ,
\]
where $W_1=K_1\left( 1-\Delta _1\right) \,$ and $W_2=K_2\left( 1-\Delta
_2\right) $.

\newpage {}

\begin{figure}[tbp]
\caption{The magnetic model for $F_2PNNNO$. (a) Uniform chains with
intramolecular ferromagnetic coupling ($J_F$) and intrachain
antiferromagnetic coupling ($J_{AF}$). The chains interact
antiferromagnetically ($J_{AF}^{^{\prime }}$). (b) The extreme limit of the
model when $J_F\to \infty $: antiferromagnetic honeycomb lattice with $S=1$.}
\label{structure}
\end{figure}

\begin{figure}[tbp]
\caption{Two-terminal graph used for renormalization purposes.}
\label{graph}
\end{figure}

\begin{figure}[tbp]
\caption{Critical inverse temperature ($K_c$) vs $C_1=J_{AF}^{^{\prime
}}/J_{AF}$ found from the RG recursion relations for different anisotropy: $%
\Delta =1.0$ (1), $\Delta =0.8$ (2), $\Delta =0.6$ (3).}
\label{critcoupl}
\end{figure}

\begin{figure}[tbp]
\caption{N\'{e}el temperature $T_N$ vs $\Delta $ phase diagram for different
values of the $C_1=J_{AF}^{^{\prime }}/J_{AF}$ ratio: $1.0$ (1); $0.5$ (2); $%
0.3$ (3). The region above (below) the critical line represents disordered
(ordered) phase. The dotted lines are guides-to-the-eyes.}
\label{critanis}
\end{figure}

\begin{figure}[tbp]
\caption{Phase diagram $T_N$ vs $\Delta $ for small $T_N$ near $\Delta _c$: $%
C_1=1.0$ and $\Delta _c\simeq 0.46$ (a), $C_1$ $=0.5$ and $\Delta _c\simeq
0.62$ (b), $C_1$ $=0.3$ and $\Delta _c\simeq 0.76$ (c).}
\label{scaling}
\end{figure}

\end{document}